\documentstyle[12pt,aaspp4,epsfig,epsf]{article}

\begin{document}
 
\title{High Energy Cosmic Rays, Gamma Rays And Neutrinos From Jetted GRBs} 
\author{Arnon Dar}
\affil{Theory Division, CERN, CH-1211 Geneva 23, Switzerland \\ 
       and \\ Department of Physics and Space Research Institute\\ 
        Israel Institute of Technology, Haifa 32000, Israel\\ } 

\begin{abstract} 
Recent observations suggest that gamma ray bursts (GRBs) and their
afterglows are produced in star formation regions in distant galaxies by
highly relativistic jets that happen to point in our direction.
Relativistic beaming collimates the emission from the highly relativistic
jets into small solid angles along the jet direction. It implies that we
are seeing only a small fraction of the events that produce GRBs.  The
observed GRB rate then requires an event rate which is comparable to the
birth rate of neutron stars (NS). The highly relativistic jets sweep up
ambient matter along their trajectories, accelerate it to cosmic ray (CR)
energies and disperse it in hot spots which they form when they stop in
the galactic halo. With an event rate comparable to the NS birth rate,
such events in our Galaxy may be the main source of Galactic cosmic rays
at all energies. Internal interactions and/or external interactions of these
jets with high column density matter and/or radiation at their production
sites or along their trajectories can produce high energy gamma rays and
neutrinos that are highly beamed along the jet direction. Jetted GRBs,
like blazars, may be much more fluent in high energy gamma rays and
neutrinos than in MeV gamma rays. But, TeV gamma rays from large cosmological
distances are unobservable because of their attenuation by
electron-positron pair creation on the intergalactic infrared background
radiation. However, high energy neutrinos from distant GRBs may be
observed with large surface/volume telescopes which are under
construction. TeV gamma rays and high energy neutrinos may also be
detected from relatively nearby GRBs by the existing moderate size
detectors, but with a much smaller rate.  
\end{abstract}

\section{The Energy Crisis Of Spherical GRBs} 
Thanks to the precise and prompt localization by the Italian-Dutch
satellite, BeppoSAX (see, e.g., Costa et al. 1997), long lived GRB
afterglows spanning the wavelength range from X-ray to radio have now been
detected in more than a dozen GRBs. They led to the redshift measurements,
z=0.69, 0.835, 3.42, 1.096, 0.966, 1.61, 1.62 of GRBs 970228 (Kulkarni et
al. 1999) 970508 (Metzger et al. 1997), 971214 (Kulkarni et al 1998),
980329, 980613 (Djorgovski et al 1999), 980703 (Djorgovski et al. 1998),
990123 (Anderson et al. 1999; Kulkarni et al. 1999), 990510 (Vreeswijk et
al. 1999), respectively, from absorption lines in their optical afterglows
and/or emission lines from their host galaxies. In addition, strong
suppression has been observed with the Hubble Space Telescope in the
spectrum of the host galaxies of GRBs 970228 and 980329 at wavelengths
below 700 nm. If it is due to absorption in the Ly$\alpha$ forest
(Fruchter 1999), then their redshifts are near
z$\sim 5$. These measured/estimated redshifts indicate that most GRBs take
place at very large cosmological distances. For instance, 
assuming a zero cosmological constant ($\Omega_\Lambda=0$), 
the luminosity  distance 
\begin{equation}
{\rm D_L= {c\over H}{2[2-\Omega_M(1-z)-(2-\Omega_M)\sqrt{1+\Omega_Mz}]
          \over \Omega_M^2}},
\end{equation} 
with the present canonical values for the 
cosmological parameters, ${\rm H=65~km~cm^{-1}~Mpc^{-1}}$ and ${\rm 
\Omega_M=0.2}$ 
(which will be assumed in this paper), 
yield ${\rm D_L\sim 5\times 10^{28}~cm}$ for z=2.
The typical observed GRB fluence (${\rm F_\gamma\sim
10^{-5~} ~erg~cm^{-2}}$) and their large distances imply enormous energy 
release in gamma rays, 
\begin{equation}
{\rm E_\gamma ={4\pi D_L^2 F_\gamma\over (1+z)}\sim 10^{53}~erg},
\end{equation}
if their energy release is isotropic 
as used to be assumed/advocated by the standard fireball
models of GRBs and GRB afterglows (e.g., Piran 1999 
and references therein). 
In particular, the large fluence ${\rm F_{\gamma}\approx 
5.1\times 10^{-4}~erg~cm^{-2}}$
(Kippen et al. 1999) and redshift z=1.61 of GRB 990123 yield 
${\rm E_{\gamma}\approx 3.4\times 10^{54}~erg}$.
Such enormous energy release in gamma rays alone, implies an ``energy 
crisis'' for
spherical GRBs (Dar 1998): The short duration and the very large energy
release in GRBs indicate that they are powered by gravitational collapse
of compact stars.  But, the energy release in such events falls short of
that required to power GRBs like 971214, 980329 and 990123, 990510
if they were
isotropic. Furthermore, all the known luminous sources of gamma rays
(quasars, radio galaxies, active galactic nuclei, accreting
binaries, pulsars, supernova explosions, supernova remnants) exhibit 
rather a modest efficiency, $\eta< 10^{-4}$, in converting gravitational,
kinetic or thermonuclear energy into gamma rays.  If GRBs have a similar
efficiency for converting the energy release from their central engine to
gamma rays, then the energy crisis is common to most GRBs. 
 
\section{Jetted GRBs}

\subsection{No Energy Crisis For Jetted GRBs}
Various authors have pointed out that the energy crisis in isotropically
emitting GRBs is avoided if GRBs are beamed into a small solid angle,
$\Delta\Omega\ll 4\pi$ such that their total energy release in gamma rays
is
\begin{equation}
{\rm E_\gamma ={\Delta\Omega D_L^2 F_\gamma\over (1+z)}}.
\end{equation}
Beaming of gamma rays from GRBs is possible if the highly relativistic
ejecta (Lorentz factor $\Gamma=1/\sqrt{1-\beta^2}\gg1 $) that produces the
GRB is beamed into a cone (conical beaming) of solid angle $\Delta\Omega
\ll 4\pi$, or if the ejecta is jetted - namely, if after initial expansion
the ejected cloud/plasmoid maintains nearly a constant cross section.
Conical beaming (e.g., Mochkovich 1993; Rhoads 1997) can
solve the `` energy crisis'' of GRBs by reducing their inferred energies
by the ratio $\Delta\Omega/ 4\pi\ll 1$. But, it suffers from other
deficiencies of isotropically emitting GRBs (Dar 1998). Jetting the
ejecta (e.g., Shaviv and Dar 1995, Dar et al. 1998, Dar 1998)
solves the energy crisis and can also explain the short time
variability of GRB light curves, the versatility of their afterglows, the
absence of a simple scaling between them and their sudden decline in some
GRBs (970508, 990123, 990510): 

The emission from a highly relativistic plasmoid which is isotropic in
its rest frame and has a power-law spectral shape, ${\rm 
F_{\nu'}=A\nu'^{-\alpha}}$,
is collimated in the lab frame to small emission angles $\theta\sim 1
/\Gamma$ relative to its direction of motion, according to
\begin{equation}
{\rm F_\nu = {2^{2+\alpha}\Gamma^{3+\alpha}\over (1+\Gamma^2\theta^2)^3}
              F_{\nu'=\nu (1+\Gamma^2\theta^2)/2\Gamma}}~~.
\end{equation}
Thus, the observed flux from a plasmoid with a typical spectral index $\alpha
\sim0.7$ that moves with a Lorentz factor $\Gamma\sim 10^3$ at an angle
$\theta<1/\Gamma$ relative to the line of sight is amplified by
approximately $\Gamma^{3+\alpha}\sim 10^{11}$. This amplification within a
solid angle $\Delta\Omega\sim\pi/\Gamma^2$ can explain why highly 
relativistic
jets with bulk motion Lorentz factors $\Gamma\sim 10^3$, total kinetic
energy ${\rm E_k\sim 10^{52}~erg}$, and conversion efficiency
$\eta>10^{-4}$ into gamma rays can produce GRBs with equivalent isotropic
energy of ${\rm E_\gamma=\eta E_k 4\pi/\Delta\Omega>4\times 10^{54}~erg}$, as
observed for GRB 990123 (Kippen et al. 1999). 

\subsection{The Beaming Angle Of GRBs}
The enormous release of energy in GRBs during a short time suggests that
they are energized by collapse of compact stars (Blinnikov 1984,
Paczynski 1986) due to mass accretion (Goodman, Dar and Nussinov 1987;
Dar et al. 1992) or phase transition (e.g., Dar 1999a). 
If GRBs are produced, e.g., by gravitational collapse of
neutron stars (NS) to quark stars (QS) 
when they
cooled and spun down sufficiently (e.g., Dar 1999a; Dar and De R\'ujula 
1999), then the GRB rate is comparable to 
the NS birth-rate. 
The NS birth-rate is estimated to be ${\rm R_{NS} \sim 0.02~y^{-1}}$ in Milky
Way like galaxies (van den Bergh and Tamman 1991). From the observed rate of 
GRBs, ${\rm R_{GRB}[UNIV]\sim 10^3~y^{-1}}$ in the whole Universe, it was
estimated that the rate of observable GRBs in Milky Way like galaxies is
${\rm R_{GRB} [MW]\sim 10^{-8}~y^{-1}}$ (e.g., Wijers et al. 1997). The 
beaming angle of GRBs therefore must satisfy
\begin{equation} 
{\rm R_{GRB}[MW]\approx 2(\Delta\Omega/4\pi) R_{NS}[MW]}
\end{equation}
where we assumed that two opposite jets are ejected in every NS collapse.
Hence, $\Delta\Omega\simeq \pi/\Gamma^2\sim \pi\times 10^{-6}$. It
implies that their bulk motion Lorentz factor is $\Gamma\sim 10^3 $. 
Such values have been inferred also from the absence of a break
due to ${\rm \gamma\gamma\rightarrow e^+e^-}$ in GRB spectra 
(e.g., Baring and Harding 1997), from the 
peak energy of GRBs and from GRB duration and substructure (e.g., 
Shaviv and Dar 1995). 
Such strong beaming implies that we observe only a very small fraction, 
$\sim 10^{-6}$,  
of the events that  produce GRBs. I will call these events cosmological
GRBs (CGRBs) if they occur in distant galaxies and 
``Galactic'' GRBs (GGRBs)  if they occur in our Milky Way (MW) galaxy.

\subsection{The Jet Energy} 

Consider gravitational collapse that leads to the birth of a pulsar
(e.g., gravitational collapse of NS to QS
due to a
phase transition of cold and highly compressed neutron matter 
to Bose condensate of diquark pairs [Dar 1999a, Dar and De R\'ujula 1999]
or the birth of a pulsar in a supernova explosion [Cen 1999])
and perhaps to the ejection of two opposite highly relativistic jets.
If momentum imbalance in the
ejection of the relativistic jets (and not asymmetric neutrino emission)
is responsible for the observed large mean velocity (Lyne and Lorimer
1994), ${\rm v\approx 450\pm 90~ km~s^{-1}}$, of slowly spinning pulsars,
then momentum conservation implies that the difference in the kinetic
energy of the jets satisfies
\begin{equation}
{\rm \Delta E_{jet} \geq cP_{ns}\sim vM_{NS}c\sim 4\times 10^{51}~erg}, 
\end{equation} 
where we used the typical observed mass of NSs, ${\rm M_{NS}\approx
1.4M_\odot}$. If ${\rm \Delta E_{jet}\ll E_{jet}}$, then the kinetic 
energy of the jets must be ${\rm E_{jet}\sim 10^{52}~erg}$ or larger.  If
$\Gamma\approx 10^3$ then the ejected jet (plasmoid) has a mass ${\rm
M_{jet}\sim 1.5\times 10^{-6}M_{NS}\sim 2.1\times 10^{-6}M_\odot\sim
0.7M_{Earth}}$. Even if only a fraction $\eta\sim 10^{-4}$ of the jet
kinetic energy is radiated in $\gamma$-rays, the inferred ``isotropic''
$\gamma$-ray emission in GRBs is ${\rm E_{isot}\simeq 4\eta \Gamma^2
E_{jet}\sim 4\times 10^{54}~erg}$, while the true $\gamma$-ray emission is
only ${\rm E_\gamma\sim 10^{48}~erg}$. 

\subsection{Jet Formation} 

Relativistic jets seem to be emitted by all astrophysical systems where
mass is accreted at a high rate from a disk onto a central compact object. 
Highly relativistic jets were observed in galactic superluminal sources,
such as the microquasars GRS 1915+105 (Mirabel and Rodriguez 1994;  Mirabel
and Rodriguez 1999) and GRO J165-40 (Tingay et al. 1995) where mass is
accreted onto a stellar black hole (BH), and in many active galactic
nuclei (AGN), where mass is accreted onto a supermassive BH. Mildly
relativistic jets from mass accretion are seen both in AGN and in star
binaries containing NS such as SS433 (e.g., Hjellming and Johnston 1988). 
The ejection of highly relativistic jets from accreting or collapsing
compact stellar objects is not well understood. Therefore their properties
must be inferred directly from observations and/or general considerations:
High-resolution radio observation resolved the narrowly collimated
relativistic jets of microquasars into blobs of plasma (plasmoids) that
are emitted in injection episodes which are correlated with sudden removal
of the accretion disk material (Rodriguez and Mirabel 1998). After initial
expansion, these plasmoids seem to retain a constant radius of $R_p\sim
10^{-3}pc$. The emission of Doppler shifted Hydrogen Ly$\alpha$ and Iron
K$\alpha$ lines from the relativistic jets of SS433 suggest that the jets
are made predominantly of normal hadronic plasma. Moreover, simultaneous
VLA radio observations and X-ray observations of the microquasar GRS
1915+105 indicate that the jet ejection episodes are correlated with
sudden removal of accretion disk material into the relativistic jets 
(Mirabel and Rodriguez 1999). 
Highly relativistic jets, probably, are also ejected in the birth or
collapse of NSs due to mass accretion or phase transition.  But,
because the accretion rates and magnetic fields involved are much larger
compared with those in quasars and microquasars, the bulk motion Lorentz
factors of these jets may be much higher, perhaps, $\Gamma \sim 10^3$ as
implied by the above consideration and GRB observations.  When these
highly relativistic jets happen to point in our direction they produce the
observed cosmological GRBs and their afterglows. They look like the
Galactic micro copies of blazer ejections and therefore will be called
``microblazers''. In fact, when the light curves and energy spectra of
blazar flares and of microquasar plasmoids are scaled according to the
Lorentz factors expected for GRB ejecta, they look quite similar to
GRBs (Dar 1999b).

The high collimation of relativistic jets over huge distances (up to tens
of pc in microquasars and up to hundreds of kpc in AGN), the confinement
of their highly relativistic particles, their emitted radiations and
observed polarizations, all indicate that the jets are highly magnetized,
probably with a strong helical magnetic field along their axis. Magnetic
fields as strong as a few tens mili Gauss in the jet rest frame have
been inferred from microquasar observations (Mirabel and Rodriguez 1999),
while hundreds Gauss were
inferred for GRB ejecta (assuming equipartition of energy between internal
kinetic and magnetic energy). The UV light and the X-rays from the jets
ionize the ISM in front of them.  The jet material and the swept-up
ionized ISM material in front of the jet can be accelerated by the Fermi
mechanism to a power-law energy distribution that extends to very high
energies, as inferred from the observed radiations from jets. The 
interactions of these high energy particles in the jet 
and/or their interactions with the external medium, produce the GRBs and 
their afterglows:

\subsection{Jet Production of Gamma Rays}

The GRB jets may consist of pure ${\rm e^+e^-}$ plasmoids or of normal
hadronic gas/plasma clouds. The GRB can be produced by electron
synchrotron emission. If the jet
consist of a single plasmoid, then individual $\gamma$-ray pulses that 
combine to form the
GRB light curve can be produced by internal
instabilities or by interaction with inhomogeneous external medium.
If the jets consist of
multiple ejections of plasmoids, then the GRB pulses may be produced when
later ejected plasmoids collide with earlier ejected plasmoids that have
slowed down by sweeping up the interstellar medium in front of them.
But, such scenarios do not seem to provide a simple explanation why
the GRB emission is peaked near $\sim$ MeV photon energy. Other GRB
emission mechanisms, however, can provide such an  explanation: 

If the highly relativistic plasmoid consists of a pure ${\rm e^+e^-}$
plasma, then inverse Compton scattering of stellar light (${\rm 
h\nu=\epsilon_{ev}\times 1~eV}$) by the plasmoid can explain the
observed typical $\gamma$ energy (${\rm \epsilon_{\gamma}\sim 4\Gamma_3^2
\epsilon_{eV} /3(1+z)~MeV}$), GRB duration ($ {\rm T\sim
R_{SFR}/ 2c\Gamma^2\sim 50~~s}$), pulse duration (${\rm t_p\sim R_p/
2c\Gamma^2\sim 150~ms}$), fluence $({\rm F_\gamma\sim
10^{-5}~erg~cm^{-2}})$, light curve and spectral evolution 
of GRBs (Shaviv and Dar 1995; Shaviv 1996; Dar 1998). For instance,
\begin{equation}
{\rm  F_\gamma \simeq{\sigma_{_T} N\epsilon_\gamma\over \Gamma m_ec^2}
{E_{jet}(1+z)\over D^2\Delta\Omega}\simeq { 10^{-5}z_2 N_{22}\Gamma_3
\epsilon_{ev} E_{52} \over D_{29}^2}~{erg\over cm^2}}  
\end{equation}
where ${\rm D=D_{29}\times 10^{29}~cm}$ is the luminosity distance of the
GRB at redshift z, ${\rm z_2=(1+z)/2}$, ${\rm N=N_{22}\times
10^{22}~cm^{-2}}$ is the column density of photons along the jet
trajectory in the star forming region, ${\rm \sigma_{_T}=0.65\times
10^{-24}~cm^{-2}}$ is the Thomson cross section, ${\rm E_{jet}=E_{52}\times
10^{52}~erg}$ and $\Gamma=\Gamma_3\times 10^3$. 

If the plasmoid consists of normal crust material of neutron stars
(Doppler-shifted ${\rm K_\alpha}$ iron line was detected from the jets of
SS433), then photoabsorption of stellar light by partially ionized heavy 
metals like iron
(Doppler-shifted to X-rays in the jet rest frame) and its reemission as
$\gamma$ rays (iron X-rays lines in the jet rest frame) yield ${\rm
\epsilon_\gamma\sim \Gamma\epsilon_x/(1+z)\sim MeV}$) in the observer
frame and
\begin{equation}
{\rm  F_\gamma \simeq{\sigma_{a} N\epsilon_\gamma\over \Gamma M_{Fe}c^2}
{E_{jet}(1+z)\over D^2\Delta\Omega}\simeq
{10^{-5} z_2\sigma_{19} N_{22}\bar{\epsilon}_x\Gamma_3E_{52} \over
D_{29}^2}~{erg\over ~cm^2}}
\end{equation}
where ${\rm \sigma_a= \sigma_{19}\times 10^{-19}~cm^2}$ is the mean
photoabsorption cross section of X-rays by partially ionized iron.  

\section{GRB Afterglows}
The afterglows of GRBs may be synchrotron emission from the decelerating
plasmoids (e.g., Chiang and Dermer 1997), and then they are highly beamed
and may exhibit  superluminal velocities 
($ {\rm c<v_\perp\leq 
\Gamma c}$) during and right after the GRB.  
The deceleration of a mildly
relativistic spherical ejecta from  NS collapse to QS may also produce
spherical supernova like light curve (many planetary nebulae, e.g., 
NGC 7009,
NGC 6826, and some SNRs (including, perhaps, SNR 1987A) show  
antiparallel jets superimposed on 
a spherical explosion).  In the rest frame of the
decelerating plasmoid, the synchrotron spectra can be modeled by
convolving the typical electron energy spectrum (${\rm E^{-p}}$ at low
energies up to some ``break energy'' where it steepens to ${\rm E^{-p-1}}$
and cuts off exponentially at some higher energy due to synchrotron losses
in magnetic acceleration)  with the synchrotron Green's function (see,
e.g., Meisenheimer et al. 1989). In the observer frame it yields spectral
intensity (Dar 1998) 
\begin{equation}
{\rm I_\nu\sim \nu^{\alpha}t^{\beta}\sim\nu^{-0.75\pm 0.25} t^{-1.25\pm 0.08}},
\end{equation}
where ${\rm \alpha =-(p-1)/2}$ and ${\rm \beta=-(p+5)/6}$ and where I
assumed  ${\rm p=2.5\pm 0.5}$ for Magnetic Fermi acceleration. This
prediction
is in agreement with observations of GRB afterglows. Moreover, the glows
of microquasar plasmoids and radio quasar jets after ejection, and of
blazar jets after flares, show the same universal behavior as observed in
GRBs afterglows. For instance, the glows of the ejected plasmoids from
GRS 1915+105 on April 16, 1994 near the source had $\alpha=-0.8\pm 0.1$ and
$\beta= - 1.3\pm 0.2$ (Rodriguez and Mirabel 1998), identical to those
observed for SS 433 (Hjellming \& Johnston 1988) and for the inner regions
of jets of some radio galaxies (e.g., Bridle \& Perley 1984). 
When the jets spread, the spectral index of
their power-law time decline changed to $\beta'=2.6\pm 0.4$ 
(e.g., Mirabel and Rodriguez 1999). Thus, their
overall time decline can be described approximately by ${\rm I\sim
t^{-\beta}/[1+(t/t_0)^{\beta'-\beta}]}$ where t$_0$ is the time when the jet
begins to spread. Indeed, such a behavior has been observed also in the 
afterglows of some GRBs (e.g., GRB 990510; Vreeswijk et al 1999).

\section {Galactic GRBs - The Main Source of Cosmic Rays?}  
According to  the current paradigm  of cosmic rays (CRs) origin, CR nuclei 
with energies
below 3 $\cdot$ 10$^{15}$ eV (the ``knee'') are accelerated in Galactic
SNRs (Ginzburg 1957) and those above $3\times 10^{18}$
eV (the ``ankle''), for which a disk origin is unlikely due to their
isotropy, in sources far beyond our Galaxy (Burbidge 1962). However,
recent observations suggest that, perhaps, SNRs are not the main 
source of Galactic cosmic rays and that the CRs above the ankle are not 
extragalactic:

- Measurable fluxes of high energy gamma rays from interactions of
cosmic ray nuclei in SNRs, as expected in models of SNR acceleration  
of CRs, were not detected from nearby SNRs (Prosch et al. 1996;
Hess et al. 1997; Buckley et al. 1998).

- The expected galactocentric gradient in the distribution of high energy
gamma rays ($>$ 100 MeV) from interactions of CRs from SNRs in the
Galactic interstellar medium is significantly larger than observed by the
EGRET detector on board CGRO (Hunter et al. 1997; Strong and Moskalenko
1998). 

- Diffusive propagation of CRs from the observed Galactic distribution of 
SNRs yield anisotropies in the distribution of CRs above 100 TeV in excess
of the observed value (Aglietta et al. 1995) by more than an order
of magnitude (Ptuskin et al. 1997).

- The absence (Takeda 1998) of the ``GZK cutoff'' in the
intensity of CRs at energies above $\sim 10^{20}$ eV due to interactions 
with the cosmic microwave radiation (Greisen 1996; 
Zatsepin \&  Kuz'min 1966) have brought into question (e.g., Hillas 1998)
their hypothesized extragalactic origin.  

Relativistic jets are efficient CRs accelerators (e.g., 
Mannheim and Biermann 1992, Dar 1998b).  
A modest fraction of the total energy
injected into the MW in jets from Galactic GRBs at a rate
similar to the NS birth/collapse rate in the MW, 
that is converted to CR energy, can supply,  $\sim 1.5\times 
10^{41}~erg~s^{-1}$, the estimated Galactic  luminosity in CRs
(Drury et al., 1989). Thus,  Dar and Plaga (1999) 
have recently proposed that
Galactic GRBs (GGRBs) are the main source of the CRs at all energies
and consequentlty no GZK cutoff is expected in the CR spectrum:

The
highly relativistic narrowly collimated jets/plasmoids from the birth or 
collapse of NSs in the disk of our Galaxy
that are emitted with ${\rm E_{jet}\sim 10^{52}~ erg}$ perpendicular to
the Galactic disk, stop only in the Galactic halo, when the rest mass
energy of the swept-up ambient material becomes $\geq$ their initial
kinetic energy. Through  the Fermi mechanism, they accelerate the 
swept-up ambient matter to CR energies and disperse it into the halo from 
the hot spots which they form when they finally stop in the Galactic halo 
(Fig.1). 
\begin{figure}[thb] \centering
\epsfig{file=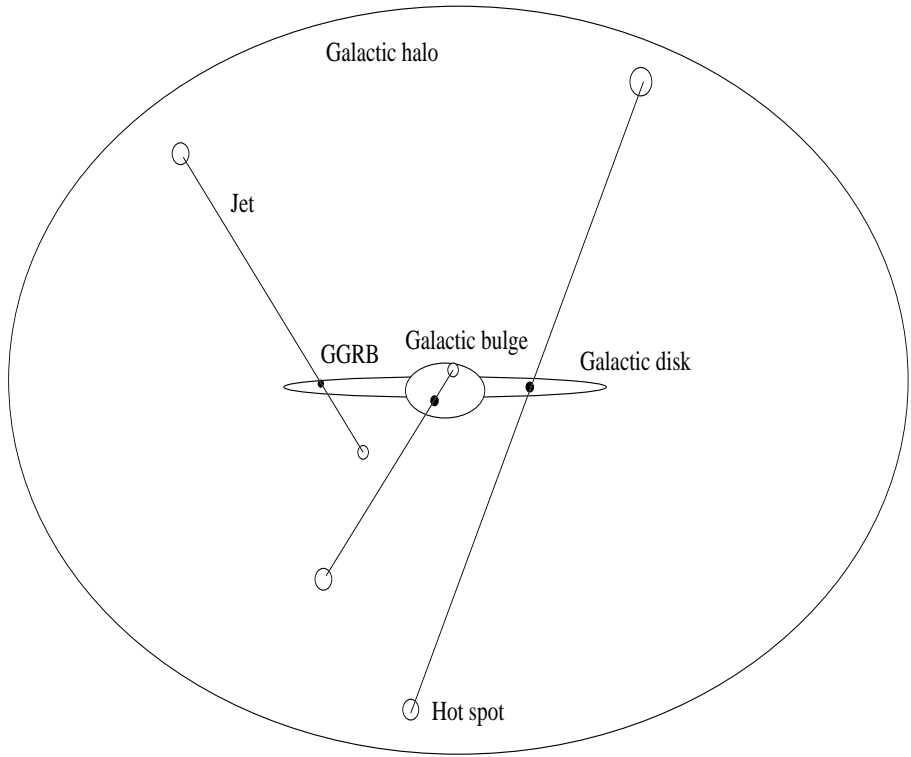,width=12cm,height=10cm,clip=,angle=0}
\caption{\label{fig1} \it A highly schematic sketch of the Dar-Plaga  
paradigm. The birth or collapse of NSs in the disk of our Galaxy leads to an
ejection of two opposite jets that produce ``hot spots'' when they stop in
an extended Galactic halo.}
\end{figure}

The typical equipartition magnetic fields in such hot spots may 
reach
${\rm B\sim (3E_{jet}/R_ p^3)^{1/2}\sim 1~ G}$. Synchrotron losses cut off
Fermi acceleration of CR nuclei with mass number A at ${\rm E\sim \Gamma
A^2Z^{-3/2}(B/G)^{-1/2}\times 10^{20}~eV.}$ Particle-escape cuts off Fermi
acceleration when the Larmor radius of the accelerated particles in the
plasmoid rest frame becomes comparable to the radius of the plasmoid,
i.e., above $ {\rm E\simeq \Gamma Z(B/G)(R_p/0.1~pc)\times 10^{20}~eV}.$
Consequently, CR with ${\rm E>Z\times
10^{20}~eV}$ can no longer be isotropized by acceleration or deflection in
hot spots with $\Gamma\sim 1$. 

Fermi acceleration in the highly relativistic jets 
from GRBs ($\Gamma\sim 10^3$)can
produce a broken power-law spectrum, ${\rm dn/dE\sim E^{-\alpha}}$, with
$\beta\sim 2.2$ below a knee around ${\rm E_{knee}\sim A~ PeV}$ and
$\beta\sim 2.5$ above this energy (Dar 1998b). Spectral indices $\alpha$
$\sim 2.2$ were obtained also in numerical simulation of relativistic
shock acceleration (e.g., Bednarz and Ostrowski 1998) . Galactic
magnetic confinement increases the density of Galactic CRs by the ratio
${\rm c\tau_{\rm h}/R_G}$ where ${\rm \tau_h(E)}$ is the mean residence
time in the halo of Galactic CRs with energy E, and ${\rm R_G\sim 50~ kpc}$
is the radius of the Galactic magnetic-confinement region. With the
standard choice for the energy dependence of the diffusion constant
(observed, e.g., in solar-system plasmas) one gets:
${\rm\tau_h\propto (E/Z)^{-0.5}}$.  Consequently, the energy spectrum of
CRs is predicted to be
\begin{equation} {\rm 
dn/dE\sim C (E/E_{ knee})^{-p}} 
\end{equation} 
with $p\simeq \alpha+0.5\simeq 2.7~ (\simeq 3)$ below (above) the
knee.  This power-law continues as long as the Galactic magnetic field
confines the CRs.

Part of the kinetic energy released by GGRBs is transported into the Galactic
halo by the jets. Assuming equipartition of this energy, without large
losses, between CR, gas and magnetic fields in the halo during the
residence time of CR there, the magnetic field strength  
B$_{\rm h}$ in the halo is
expected to be comparable to that of the disk ${\rm B_ h \sim
(2L_{\rm MW}[CR]\tau_ h /R_ h^3)^{1/2}\simeq 3~\mu G}$
where ${\rm \tau_ h\sim 5\times
10^9~ y}$ is the mean residence time of the bulk of the CRs in the Galactic
halo. Cosmic rays with Larmor radius larger than 
the coherence length $\lambda$  of the halo magnetic fields, i.e., with
energy above  
\begin{equation} {\rm
E_{ankle}\sim 3\times 10^{18}(ZB_h/3\mu G)(\lambda/kpc)~eV}, 
\end{equation}
escape Galactic trapping. Thus, the
CR ankle is explained as the energy where the mean residence time
${\rm \tau_h(E)}$ of CRs becomes comparable to the free escape time from
the halo
$\tau_{\rm free}\sim 1.6(R_{\rm h}/50~{\rm kpc})\times 10^5~{\rm years}$.
Therefore, the spectrum of CRs with energies above the ankle, that do not 
suffer Galactic magnetic trapping, is the CRs spectrum produced by the 
jet, i.e., 
\begin{equation} 
{\rm
dn/dE\sim C (E_{ankle}/E_{knee})^{-3}(E/E_{ankle})^{-2.5};~
E> E_{ankle}}.
\end{equation}
Eqs. 10-12 describe well the overall CR energy spectrum.
\begin{figure}[thb]
\centering \epsfig{file=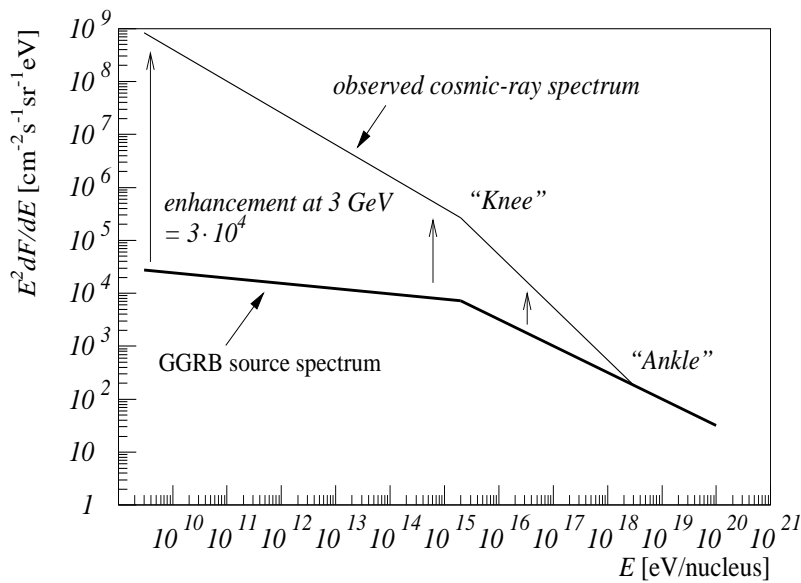,width=12cm,height=10cm,clip=,angle=0}
\caption{\label{fig2} \it The observed flux of cosmic rays
(thin line) as a function of primary energy E 
is well described by a power law that changes
its slope sharply at only two
energies the ``knee'' and the ``ankle''.
At energies below the ankle
it is enhanced (by a factor (E/E$_{ankle}$)$^{-0.5}$)  over the GGRB source spectrum
(thick line, a power law with differential power law index
of -2.2 below the knee and $\simeq$ -2.5 above it)
by way of trapping in the
Galactic halo magnetic fields. } 
\end{figure}

\section{High Energy Gamma Rays} 

The observed radiations from blazars, microquasars and GRBs 
indicate that their highly relativistic jets contain high energy 
charged particles with a power-law energy distribution 
${\rm dn_p/dE\sim AE^{-\alpha}}$ with $\alpha\sim 2.2$, that extends
to very high energies. Such distributions can be formed in the highly
relativistic jets through Fermi acceleration of swept up material. 
This power-law distribution of energetic protons is boosted and beamed into a
solid angle $\Delta\Omega\approx \pi/\Gamma^2$ in
the lab frame. GRB afterglows suggest
that GRBs are produced in star formation regions, probably molecular
clouds. The typical column density of such clouds is $N_p=10^{24\pm 1}
cm^{-2}$. The clouds must be highly ionized 
along the line of sight to the GRB by the
enormous fluxes of beamed X-rays and UV radiations from the GRB.
Interaction of the highly relativistic GRB jets with the high column
density of this ionized gas (or with diffuse matter at their production
sites) can produce high energy gamma rays through ${\rm pp\rightarrow
\pi^0(\eta^0)X}$;  $\pi^0(\eta^0)\rightarrow 2\gamma$. The cross section for 
inclusive
production of high energy $\gamma$-rays with a small transverse momentum,
${\rm cp_{T}=E_{T}<1~ GeV}$ in pp collisions (e.g., Neuhoffer et al. 1971; 
Boggild and Ferbel 1974; Ferbel and Molzon 1984) is well represented by
\begin{equation} 
{\rm {E\over \sigma_{in}} {d^3\sigma\over d^2p_{T}
dE_\gamma}\approx (1/2\pi p_{T}) e^{-E_{T}/E_0}~f(x)}, 
\end{equation} 
where $E$ is the incident proton energy, ${\rm \sigma_{in} 
\approx 35~mb}$ is 
the pp total inelastic cross section at TeV energies, $E_0\approx 0.16~
GeV$ and ${\rm f(x)\sim (1-x)^3/\sqrt{x}}$ is a function only of the Feynman
variable ${\rm x=E_\gamma/E}$, and not of the separate values of the 
energies of
the incident proton and the produced $\gamma$-ray (Feynman scaling).
The exponential
dependence on ${\rm E_{T}}$ beams the $\gamma$-ray production into 
${\rm \theta <
E_{T}/E\sim 0.17/\Gamma}$ along the incident proton direction. When
integrated over transverse momentum the inclusive cross section becomes
${\rm \sigma_{in}^{-1}d\sigma/ dx\approx f(x)}.$ 
If the incident protons have a
power-law energy spectrum, ${\rm dn_p/dE\approx AE^{-\alpha}}$, then, 
because of
Feynman scaling, the produced $\gamma$-rays have the same power-law
spectrum:  
\begin{equation} 
{\rm {dn_\gamma\over d E_\gamma}
     \approx N_p \sigma_{in} \int_{E_\gamma}^{\infty} {dn_p\over dE}
      {d\sigma\over dE_\gamma}dE 
      \approx N_p \sigma_{in}g_{p\gamma}AE_\gamma^{-\alpha}},  
\end{equation} 
where ${\rm N_p}$ is the column density of the
target and ${\rm g_{p\gamma}=\int_0^1x^{\alpha-1}f(x)dx\approx 0.092 
}$ for $\alpha\approx 2.2$. Consequently, 
the collimated flux of high energy
gamma rays produced by a GRB jet with initial kinetic energy 
${\rm E=E_{52}\times 10^{52}~erg}$
that propagates through a molecular cloud of typical
column density ${\rm N_p=N_{23}\times 10^{23}~cm^{-2}}$ 
is given by
\begin{equation}
{\rm {dn_\gamma\over dE} \approx {6\times 10^{-6}
 E_{52}N_{23}\Gamma_3^2\over D_{29}^2}(1+z)^{2-\alpha}
\left[ {E\over TeV}\right]^{-2.2}e^{-\tau(z,E)}~cm^{-2}~ TeV^{-1}}, 
\end{equation}
where ${\rm D(z)=D_{29}\times 10^{29}~cm}$, is the luminosity distance to 
the GRB and
${\rm \tau(z,E)}$ is the optical depth to the GRB (redshift z) at energy 
E. The
fluxes of high energy gamma rays from GRBs at $z\sim 2$ are attenuated
strongly ($\tau>1$) for E$>20~$GEV. For not very distant GRBs, e.g.,
z$< 0.5$, the gamma ray flux is not attenuated strongly at energy
${\rm E< 100~GeV}$ (Salomon and Stecker, 1998). GRBs with z$<0.1$
(${\rm D_L< 500 Mpc}$) can be 
visible in TeV gamma rays. But, their expected rate is only
\begin{equation}
{\rm R_{GRB}(z<0.1) \sim R_{GRB}[L_*]\rho_{_L}V_c(z<0.1)\sim 0.1~y^{-1}},
\end{equation}
where ${\rm R_{GRB}[L_*]\approx 2\times 10^{-8}~y^{-1}}$ is the estimated
mean rate of GRBs in ${\rm L_*\approx 10^{10}L_\odot}$ galaxy for z$<0.1$,
${\rm \rho_{_L}\sim 1.2\times 10^8L_\odot~Mpc^{-3}}$ is the luminosity 
density in the local Universe and ${\rm V_c(z<0.1)\approx 5\times 10^8Mpc^3}$
is the comoving volume within z$<0.1$. GeV gamma rays from 3 very luminous
GRBs have been reported by the EGRET detector on board CGRO
(Dingus et al 1994; Dingus 1995), consistent with the above predictions.
However, they could have also been 
produced by 
inverse Compton scattering of GRB photons from energetic electrons in 
the GRB jets. Only the 
detection of high energy
neutrinos from GRBs can establish the hadronic production origin of 
high energy photons from GRBs, i.e., the hadronic nature of the GRB jets.

\section{High Energy Neutrinos From GRBs}
Hadronic production of photons in diffuse targets is accompanied by
neutrino emission mainly through hadronic production of mesons that 
decay into neutrinos, e.g., 
${\rm pp\rightarrow\pi^{\pm}\rightarrow
\mu^{\pm}\nu_\mu}$ ; ${\rm \mu^{\pm} \rightarrow e^{\pm}\nu_\mu\nu_e}$.
Analytical calculations (e.g., Dar 1983, Dar 1984, Lipari 1993)  
show that  a proton power-law spectrum, ${\rm dn_p/dE= AE^{-\alpha}}$ with a
power index  $\alpha\sim 2.2$, generates power-law spectra 
of $\gamma$-rays and $\nu_\mu$'s that satisfy approximately,
${\rm dn_\nu/dE\approx 0.80 dn_\gamma/dE}$ (Dar and Shaviv, 1996). 
Consequently, \begin{equation}
{\rm {dn_\nu\over dE} \approx {5\times 10^{-6}
E_{52}N_{23}\Gamma_3^2\over D_{29}^2}(1+z)^{2-\alpha}
\left[ {E\over TeV}\right]^{-2.2}~cm^{-2}~ TeV^{-1}}. 
\end{equation}
Thus, we predict that the high energy $\gamma$-ray emission from 
GRBs is
accompanied by emission of high energy neutrinos with similar fluxes,
light curves and energy spectra. The expected number of $\nu_\mu$ events 
from a GRB in a deep underwater/ice  $\nu_\mu$ telescope is  
\begin{equation}
{\rm N_{events}\approx S N_A \int
R_\mu{d\sigma_{\nu\mu}\over dE_\mu}{dn_\nu\over dE}dE_\mu dE}
\end{equation}
where S is the effective surface area of the telescope, ${\rm N_A}$ is 
Avogadro's 
number, $\sigma_{\nu\mu}$ is the inclusive cross section for ${\rm \nu_\mu p
\rightarrow \mu X}$, and ${\rm R_\mu}$ is the range (in ${\rm 
gm~cm^{-2}}$) of muons with energy $E_\mu$ in water/ice. 
The number of events is not sensitive to the detector energy threshold, 
if it is below 300 GeV where both the muon range and production cross 
section increase linearly with energy and yield a 
detection probability in ice/water, which increases like
${\rm \approx 10^{-6}(E/TeV)^2}$ below 300 GeV. 
Using the neutrino cross sections that were calculated by 
Gandhi et al (1998) and neglecting detector threshold 
effects and neutrino attenuation in Earth (which becomes 
important only above 100 TeV), 
we predict that the number of neutrino events in deep 
underground ice/water neutrino telescopes per 1  km$^2$ is  
\begin{equation}
{\rm N_{events} \approx 1.3\times (1+z)^{-0.2}{E_{52}N_{23}\Gamma_3^2\over 
D_{29}^2}~km^{-2}~}. 
\end{equation}
Thus, a relatively nearby GRB (z=0.5) may generate  
${\rm \sim 40E_{52}N_{23}}$ upgoing muon events in underwater/ice telescope 
per 1 km$^2$ area and only ${\rm \sim 1E_{52}N_{23}}$ events if it is at 
z=2.   The expected time length of these neutrino bursts from GRBs
is typically, ${\rm t\approx R_{cl}/c\Gamma^2
\sim R_{10}\times 10^3~s}$, where ${\rm R_{cl}=10R_{10}~pc}$ is the size of 
the molecular cloud. Such events can be distinguished from the atmospheric
neutrino background by their directional and time coincidence
with the GRBs and establish the hadronic nature of the relativistic 
jets from GRBs. 

Unlike the neutrino bursts from nearby supernova explosions, the arrival
times of $\nu's$ from GRBs, which are spread over ${\rm t\geq t_3\times
10^3~s}$, yield only poor limits on neutrino masses and lifetimes: ${\rm
m_\nu c^2>\sqrt{2t/T}E_{\nu}\sim \sqrt{t_3/T_{10}}[E_\nu/Tev]\times 
10^5~eV}$,
where ${\rm T_{10}}$ is the GRB lookback time in units of 10Gy. This limit
cannot compete with the cosmological limit, ${\rm \sum m_{nu}c^2<
94\Omega_Mh^2~eV \approx 8~eV}$, for long lived neutrinos. The neutrino
arrival times can be used, however, to improve the limit from Supernova
1987A (LoSecco 1987) on the equivalence principle of General Relativity.

\begin{acknowledgements} Results presented here are based on
an ongoing collaboration with A. De R\`ujula and R. Plaga.
This research was supported by the fund for the promotion of research at 
the Technion.
\end{acknowledgements}

 \end{document}